\documentclass[letterpaper, 10 pt, conference]{ieeeconf}  

\IEEEoverridecommandlockouts          
\usepackage{graphics}
\usepackage{subcaption}
\usepackage{epsfig} 
\usepackage{amsmath}
\usepackage{amssymb}
\usepackage{ulem}
\usepackage{siunitx}
\usepackage{soul}
\usepackage[hidelinks]{hyperref}
\usepackage{textcomp}
\usepackage{mathtools}
\usepackage[dvipsnames]{xcolor}
\usepackage{multirow}

\newenvironment{problem*}{%
  \par\textit{Problem:} 
}{\par} 

\newenvironment{theorem*}{%
  \par\textit{Theorem:} 
}{\par} 

\usepackage[acronym,nonumberlist,nopostdot,nogroupskip]{glossaries}
\newacronym{kf}{KF}{Kalman filter}
\newacronym{ekf}{EKF}{extended Kalman filter}
\newacronym{eqf}{EqF}{equivariant filter}
\newacronym{ltv}{LTV}{linear time-varying}
\newacronym{slam}{SLAM}{simultaneous localization and mapping}
\newacronym{iekf}{IEKF}{invariant extended Kalman filter}
\newacronym{imu}{IMU}{inertial measurement unit}
\newacronym{fov}{FoV}{field of view}
\newacronym{ibvs}{IBVS}{image-based visual servoing}
\newacronym{mle}{MLE}{maximum likelihood estimator}

\newcommand{\tr}[1]{{\ensuremath{\textrm{#1}}}}   
\newcommand{\mbb}[1]{{\ensuremath{\mathbb{#1}}}}  
\newcommand{\mf}[1]{{\ensuremath{\mathfrak{#1}}}}  
\newcommand{\mc}[1]{{\ensuremath{\mathcal{#1}}}}  
\newcommand{\mr}[1]{{\ensuremath{\mathrm{#1}}}}   
\newcommand{\mb}[1]{{\ensuremath{\mathbf{#1}}}}   
\newcommand{\bs}[1]{\ensuremath{\boldsymbol{#1}}} 

\newcommand{\mring}[1]{\mathring{#1}}
\newcommand{\T}[0]{\top}

\DeclarePairedDelimiterX{\norm}[1]{\lVert}{\rVert}{#1}
\DeclareMathAlphabet{\mathpzc}{OT1}{pzc}{m}{it} 

\usepackage{todonotes}

\title{\LARGE \bf
Equivariant Observer for Bearing Estimation\\with Linear and Angular Velocity Inputs
}

\author{Gil Serrano, Marcelo Jacinto, Bruno J. Guerreiro, and Rita Cunha%
\thanks{The work of G. Serrano and M. Jacinto was supported by the PhD Grants from MIT Portugal and Funda\c{c}{\~a}o para a Ci{\^e}ncia e a Tecnologia
(FCT) [DOI: \href{https://doi.org/10.54499/PRT/BD/154275/2022}{10.54499/PRT/BD/154275/2022} and \href{https://doi.org/10.54499/2022.09587.BD}{10.54499/2022.09587.BD}]. This work was also supported by FCT, Portugal through LARSyS [DOI: \href{https://doi.org/10.54499/LA/P/0083/2020}{10.54499/LA/P/0083/2020}].}
\thanks{The authors are with the Institute for Systems and Robotics, Laboratory of Robotics and Engineering Systems, Instituto Superior Técnico, University of Lisbon, Portugal. B. J. Guerreiro is also with CTS/Uninova and LASI, School of Science and Technology, NOVA University Lisbon, Caparica, Portugal.}
\thanks{E-mails: {gil.serrano@tecnico.ulisboa.pt}, {bj.guerreiro@fct.unl.pt}, {\{mjacinto, rita\}@isr.tecnico.ulisboa.pt}}}

\begin{document}

\maketitle
\thispagestyle{empty}
\pagestyle{empty}


\begin{abstract}
This work addresses the problem of designing an equivariant observer for a first order dynamical system on the unit-sphere. Building upon the established case of unit bearing vector dynamics with angular velocity inputs, we introduce an additional linear velocity input projected onto the unit-sphere tangent space. This extended formulation is particularly useful in image-based visual servoing scenarios where stable bearing estimates are required and the relative velocity between the vehicle and target features must be accounted for. Leveraging lifted kinematics to the Special Orthogonal group, we design an observer for the bearing vector and prove its almost global asymptotic stability. Additionally, we demonstrate how the equivariant observer can be expressed in the original state manifold. Numerical simulation results validate the effectiveness of the proposed algorithm.
\end{abstract}





\section{Introduction}
\label{sec:introduction}
Nonlinear observer design for systems evolving on spheres has gained significant attention due to its wide range of practical applications, including attitude estimation\;\cite{van_goor_equivariant_filter_2023}, relative target localization\;\cite{batista_6160264}, \gls{ibvs} control strategies\;\cite{ibvs_florent_le_bras}, and \gls{slam}\;\cite{LOURENCO201861}. When relying solely on bearing measurements for relative localization applications, several estimation frameworks have been proposed in the literature. Early approaches decomposed the bearing vector using polar coordinates and made use of an \gls{ekf} or a \gls{mle} to stabilize the
linearized estimation error and estimate the position and velocity of a target\;\cite{FARINA199961}. To solve an analogous source localization problem, Batista et al.\;\cite{Bastista_source_localization} represent the relative direction information as a unit bearing vector and transform a nonlinear system into an \gls{ltv} system, enabling the use of a \gls{kf} solution. This approach was later extended for \gls{slam} scenarios involving multi-bearing measurements\;\cite{LOURENCO201861,Batista_multi_bearing_measurements}. Nonlinear adaptive observer design techniques have also been proposed by Le Bras et al.\;\cite{pe_lebras_2027_springer} and Vasconcelos et al.\;\cite{VASCONCELOS2010155} to estimate the position and velocity of an agent relative to a set of landmarks using bearing vector measurement.

Recent advancements in filter and observer designs have focused on more sophisticated techniques that exploit the geometric symmetries in system dynamics. Bonnabel et al.\;\cite{barrau_invariant_2017} introduced the concept of the \gls{iekf}, which leverages the geometric structure inherent in the state space and the system dynamics by representing them on Lie groups. These ideas have also found successful applications in attitude estimation and \gls{slam} problems\;\cite{4434662, invariant_kalman_filter}. Mahony et al. also explored the idea of designing nonlinear observers for kinematic systems with symmetry properties\;\cite{complementary_mahony}, and later also proposed an \gls{eqf} framework and applied it to the problem of estimating the bearing of a fixed direction in an inertial frame with respect to a rotating frame with angular velocity inputs\;\cite{van_goor_equivariant_filter_2023}.

Within \gls{ibvs} control frameworks, it is commonplace to combine angular velocity measurements provided by an \gls{imu}, with tracked image features provided by a camera sensor directly into the control strategy. These features can also be projected onto a virtual unit sphere and transformed into bearing-vectors \cite{ibvs_florent_le_bras, 11186936}. Additionally, the optical flow provided by the camera sensor can provide a scaled relative linear velocity measurement between the vehicle and the target features. However, while utilizing direct measurements in feedback control can appear a straightforward approach, in practice it can lead to robustness issues due to noise and outliers in the measurements. Consequently, nonlinear observers are often necessary to provide stable bearing vector estimates to the control system.

Inspired by practical challenges of \gls{ibvs} applications, in this work we propose an observer for a first order dynamical system that evolves on the unit-sphere. Our main contribution is the design of an equivariant observer for determining the bearing to a target image feature in a translating and rotating body frame. Building upon the framework established by van Goor et al.\;{\cite{van_goor_equivariant_filter_2023}}, initially developed to provide stable bearing direction estimates with respect to a rotating frame with only angular velocity inputs provided by an \gls{imu}, we extend the first-order system to incorporate distance-normalized linear velocity inputs lying on the unit-sphere's tangent space, provided by optical flow. By defining the necessary maps, according to the underlying group symmetry, we derive an equivariant lift and design a correction term that ensures almost global asymptotical stability of the observer. Then, we also demonstrate how an equivalent observer can be recovered directly on the unit-sphere manifold. Numerical simulations are presented to evaluate the performance of the proposed method and demonstrate its robustness to noise and outlier measurements.

The remainder of this paper is organized as follows: Section\,\ref{sec:notation} provides the notation adopted, mathematical properties and definitions; 
Section\,\ref{sec:problem_statement} introduces the bearing estimation problem and describes the system under consideration; 
Section\,\ref{sec:equivariant_system_formulation} presents the formulation needed to derive an equivariant observer; 
Section\,\ref{sec:equivariant_observer_design} details the observer design process; 
Section\,\ref{sec:numerical_results} presents the simulation results that demonstrate the performance of the proposed method. Section\,\ref{sec:conclusion} offers concluding remarks.


\section{Preliminaries and Notation} \label{sec:notation}

In this section, we introduce the notation adopted along with the definitions and properties used throughout this work.

\subsection{Special Orthogonal Group and Related Properties}

The Special Orthogonal group, $SO(3)$, is a matrix Lie group that represents the
set of all rotations in three-dimensional space. The group and its Lie-algebra, $\mf{so}(3)$, are defined as
\begin{align} 
    SO(3) &\coloneqq \left\{ \mb{R} \in \mbb{R}^{3 \times 3} \mid \mb{R} \mb{R}^{\top} = \mb{I},\, \text{det}(\mb{R}) = 1 \right\},\\ 
    \mf{so}(3) &\coloneqq \left\{ \mb{U} \in \mbb{R}^{3 \times 3} \mid \mb{U} = -\mb{U}^{\top} \right\}. 
\end{align}
The identity element is the ${3 \times 3}$ identity matrix, denoted by $\mb{I}$, and the group operation is simply matrix multiplication.

The skew-map ${\bs{S}(\cdot): \mbb{R}^3 \to \mf{so}(3)}$ maps a vector ${\mb{a} \in \mbb{R}^3}$ to a skew-symmetric matrix, such that ${\bs{S}(\mb{a}) \mb{b} = \mb{a} \times \mb{b}}$, for all ${\mb{b} \in \mbb{R}^3}$. It is an isomorphism between the Euclidean space $\mbb{R}^{3}$ and the Lie-algebra $\mf{so}(3)$. Consider an element of the Lie-algebra ${\mb{U} \in \mf{so}(3)}$. The adjoint map ${\text{Ad}_{\mb{R}}: \mf{so}(3) \to \mf{so}(3)}$, for ${\mb{R} \in SO(3)}$, is defined as ${\text{Ad}_{\mb{R}}\mb{U} = \mb{R}\mb{U}\mb{R}^{\top}}$. For any rotation matrix ${\mb{R}\in SO(3)}$ and vector ${\mb{u}\in\mbb{R}^{3}}$, the following property holds:
\begin{equation}
\bs{S}(\mb{R}\mb{u}) = \mb{R}\bs{S}(\mb{u})\mb{R}^{\top}\, = \text{Ad}_{\mb{R}}\bs{S}(\mb{u}).
\label{eq:skew_rotation} 
\end{equation}

\subsection{Smooth Manifolds and the 2-Sphere}

A smooth manifold of dimension $m$ is a topological space that is locally homeomorphic to an Euclidean space, $\mbb{R}^m$. For a smooth manifold $\mc{M}$, the tangent space at a point ${\bs{\xi} \in \mc{M}}$, denoted by $T_{\bs{\xi}}\mc{M}$, is a vector space that consists of all tangent vectors at that point. 

Consider a differentiable function $h$ between two smooth manifolds $\mc{M}$ and $\mc{N}$, i.e., ${h: \mc{M} \to \mc{N}}$. The differential of the function $h$, with respect to the argument $\bs{\xi}$ is denoted by $D h(\bs{\xi})$. When evaluated at a point ${\bs{\xi}^{\prime} \in \mc{M}}$, the differential denoted by $D h(\bs{\xi})_{|\bs{\xi}={\bs{\xi}^\prime}}$, along a direction ${\bs{v}\in T_{\bs{\xi}^\prime}\mc{M}}$, is defined as a linear map from the tangent space $T_{\bs{\xi}^\prime}\mc{M}$ to the tangent space $T_{h(\bs{\xi}^\prime)}\mc{N}$, i.e., 
\begin{equation}
    \begin{split}
        D h(\bs{\xi})_{|\bs{\xi}={\bs{\xi}^\prime}}: T_{\bs{\xi}^\prime}\mc{M} &\to T_{h(\bs{\xi}^\prime)}\mc{N}\\
        \bs{v} &\mapsto D h(\bs{\xi})_{|\bs{\xi}={\bs{\xi}^\prime}}[\bs{v}].
    \end{split}
    \label{eq:differential_map}
\end{equation}
The differential is given by
\begin{equation}
    D h(\bs{\xi})_{|\bs{\xi}={\bs{\xi}^\prime}}[\bs{v}] \coloneqq \lim_{t\to0} \frac{h(\bs{\xi} + t\bs{v}) - h(\bs{\xi})}{t} \Big|_{\bs{\xi}=\bs{\xi^{\prime}}}\,.  
    \label{eq:differential_definition}
\end{equation}

The 2-sphere (or unit-sphere), $\mc{S}^{2}$, is a smooth manifold defined as the set of points in $\mbb{R}^{3}$ that are at a unit distance from the origin, that is, ${\mc{S}^{2} := \{\mb{y} \in \mbb{R}^{3} \mid \norm{\mb{y}} = 1\}}$. The tangent space at a point ${\mb{y} \in \mc{S}^{2}}$ is given by ${T_{\mb{y}}\mc{S}^{2} = \{\mb{z} \in \mbb{R}^{3} \mid \mb{y}^{\top}\mb{z} = 0\}}$, i.e., the set of vectors in $\mbb{R}^{3}$ that are orthogonal to $\mb{y}$.

The projection operator $\bs{\Pi}_{\mb{y}}$ projects a vector ${\mb{x} \in \mbb{R}^3}$ onto the plane orthogonal to ${\mb{y} \in \mc{S}^2}$. It is given by ${\bs{\Pi}_{\mb{y}}:=\mb{I}_{3}-\mb{y}\mb{y}^\top}$ and the projection is $\bs{\Pi}_{\mb{y}}\mb{x}$. The projection operator is related to the skew-map by ${\bs{\Pi}_{\mb{y}} = - \big(\bs{S}(\mb{y})\big)^{2}}$. 

\subsection{Invariance, Equivariance and Group Actions}

Consider a system described by 
\begin{equation}
    \dot{\bs{\xi}} = f(\bs{\xi},\mb{u}), \quad \mb{y} = h(\bs{\xi}),
    \label{eq:general_system}
\end{equation}
where the state $\bs{\xi}$ belongs to the state manifold $\mc{M}$, the input $\mb{u}$ to the input manifold $\mc{U}$, and the output $\mb{y}$ to the output manifold $\mc{Y}$. Assume as well that there is a Lie group $G$ that acts transitively on $\mc{M}$ and $\mc{U}$ through the right actions $\phi: G \times \mc{M} \to \mc{M}$ and $\psi: G \times \mc{U} \to \mc{U}$, respectively, and ${\mb{X}\in G}$ is an element of the Lie group.

The system is said to be \textit{invariant}, with respect to the group action $\phi$, if the dynamics remain unchanged under the transformation induced by $\phi$, i.e., for all ${\mb{X} \in G}$, ${\bs{\xi} \in \mc{M}}$, and ${\mb{u} \in \mc{U}}$, the following condition is verified:
\begin{equation}
D\phi_{\mb{X}}(\bs{\xi})[f(\bs{\xi}, \mb{u})] = f(\phi_{\mb{X}}(\bs{\xi}), \mb{u}),
\label{eq:invariant_system_condition}
\end{equation}
where the notation $\phi_{\mb{X}}(\bs{\xi})$ indicates that the argument $\mb{X}$ is fixed and $\phi$ is evaluated with respect to the argument $\bs{\xi}$.

On the other hand, the system is said to be \textit{equivariant}, with respect to the group actions $\phi$ and $\psi$, if the dynamics change in a structured manner under the transformations induced by $\phi$ and $\psi$\;\cite{mahony_observer_2022}, i.e., for all ${\mb{X} \in G}$, ${\bs{\xi} \in \mc{M}}$, and ${\mb{u} \in \mc{U}}$, the following condition holds:
\begin{equation}
D\phi_{\mb{X}}(\bs{\xi})[f(\bs{\xi}, \mb{u})] = f(\phi_{\mb{X}}(\bs{\xi}), \psi_{\mb{X}}(\mb{u})).
\label{eq:equivariant_system_condition}
\end{equation}
From the definitions, it follows that invariant systems are a subclass of equivariant systems, where the input transformation is trivial.

We also introduce the notion of the stabilizer subgroup of an action $\phi$ at a point ${\bs{\xi} \in \mc{M}}$, which is defined as the set
\begin{equation}
    \tr{stab}_{\phi}(\bs{\xi}) \coloneqq \{ \mb{X} \in G \mid \phi(\mb{X}, \bs{\xi}) = \bs{\xi} \}.
\end{equation}
This subgroup contains all the elements of the Lie group\;$G$ that leave the point $\bs{\xi}$ unaltered under the action $\phi$, as is the case of the identity element of the group.

\section{Problem Statement}
\label{sec:problem_statement}
Let ${\mb{p}_{\mathrm{B}} \in \mathbb{R}^{3}}$ and ${\mb{R} \in SO(3)}$ denote the position and attitude of a vehicle's body frame $\{\mathcal{B}\}$, with respect to an inertial frame $\{\mathcal{I}\}$, expressed in $\{\mathcal{I}\}$. Analogously, let ${\mb{p}_{\mathrm{T}} \in \mathbb{R}^{3}}$ the position of a target frame $\{\mathcal{T}\}$, expressed in $\{\mathcal{I}\}$ and ${\mb{p} \coloneqq \mb{p}_{\mathrm{T}} - \mb{p}_{\mathrm{B}}}$ denote the relative position between the vehicle and the target, expressed in $\{\mathcal{I}\}$. 

\begin{figure}
    \centering
    \includegraphics[width=\columnwidth]{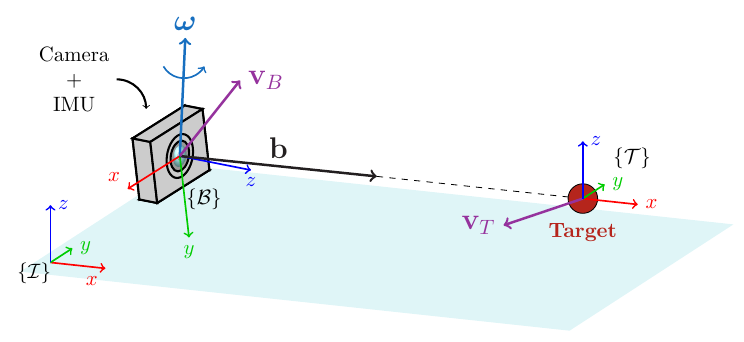}
    \caption{Bearing system: a camera translating and rotating in space, while tracking a target point also moving in space.}
    \label{fig:bearing_problem}
\end{figure}

Consider the \gls{ibvs} setup presented in Fig.\;\ref{fig:bearing_problem}, where the vehicle is equipped with a monocular camera aligned with $\{\mathcal{B}\}$, capable of measuring a bearing vector pointing towards the target, given by
\begin{equation}
    \mb{b} \coloneqq \mb{R}^{\top}\frac{\mb{p}}{\|\mb{p}\|} \in \mc{S}^{2}.
\end{equation}
Taking the time-derivative of the bearing vector, yields
\begin{equation}
	\dot{\mb{b}} = -\bs{S}(\bs{\omega})\mb{b} + \mb{\bar{v}},\quad \mb{\bar{v}} \coloneqq \frac{1}{\|\mb{p}\|}\bs{\Pi}_{\mb{b}}\mb{R}^{\top}\dot{\mb{p}},
	\label{eq:example_system}
\end{equation}
where $\bs{\omega} \in \mbb{R}^3$ is an angular velocity provided by an \gls{imu} and $\mb{\bar{v}} \in \mathbb{R}^{3}$ is a scaled linear velocity that lives in the tangent space of the unit sphere $T_{\mb{b}}\mc{S}^{2}$, obtained via optical flow using the monocular sensor.

\begin{problem*}
    Consider the dynamical system given by \eqref{eq:example_system}, where a monocular sensor provides noisy bearing vector measurements, and the linear and angular velocity inputs are provided by an \gls{imu} and optical flow, respectively. Design an equivariant observer capable of providing filtered bearing vector estimates that can be used by an \gls{ibvs} control system.
\end{problem*}

\section{Equivariant System Formulation}
\label{sec:equivariant_system_formulation}

In an equivariant observer, we want to leverage the symmetries that are inherent to the problem by lifting its dynamics to a Lie group $G$. With the system dynamics on the group, we can take advantage of the group operations and derive an observer whose estimates can be translated back to the state manifold.

\subsection{Symmetry of the Bearing System}
In this work, the state of the system lives in the 2-sphere, that is, the system evolves on the manifold ${\mc{M}=\mc{S}^{2}}$, the input manifold is ${\mc{U}=\mbb{R}^{3} \times \mbb{R}^{3}}$, and the output manifold is ${\mc{Y}=\mc{S}^{2}}$. Using the notation defined, we have that ${\bs{\xi} = \mb{b}}$, ${\mb{u} = (\bs{\omega}, \mb{\bar{v}})}$, and ${\mb{y} = h(\bs{\xi}) = \bs{\xi}}$. The dynamics of the system are given by
\begin{align}
    f(\bs{\xi},\mb{u}) &= -\bs{S}(\bs{\omega})\bs{\xi} + \mb{\bar{v}}.
\end{align}
Similarly to the well-studied problem where ${\mb{\bar{v}} = \mb{0}}$, we note that the symmetry group inherent to this system is the Lie group ${G:=SO(3)}$, with Lie-algebra ${\mf{g}:=\mf{so}(3)}$\;\cite{mahony_observer_2022}.

\subsection{Transitive Right Group Action}
The next step is to define a map $\phi: G \times \mc{M} \to \mc{M}$, in which the variable on the group acts on the variable on the manifold, translating it from one point on the manifold to another by performing a right group action on the manifold. For the system at hand, this function is
\begin{equation}
	\phi(\mb{X},\bs{\xi}) = \mb{X}^{\top}\bs{\xi},
    \label{eq:state_action}
\end{equation}
with $\mb{X} \in SO(3)$, i.e. $\mb{X}$ is a rotation matrix. For proof that $\phi$ is a transitive right group action, see Appendix\,\ref{app:right_action_phi_proof}.

\subsection{Coordinate System's Origin}
We need to define an origin $\mring{\bs{\xi}}$ for our coordinate system, in order to translate the group state to the state on the manifold. The origin can arbitrarily be defined in the manifold $\mc{M}$ as $\mring{\bs{\xi}} = \mb{e}_{3} = [0 \,0 \, 1]^{\top}$, for example. Then, $\bs{\xi}$ is given by the transpose of the third row of the rotation matrix, according to
\begin{equation}
	\bs{\xi} = \phi(\mb{X},\mring{\bs{\xi}}) = \mb{X}^{\top}\mring{\bs{\xi}} = \mb{X}^{\top}\mb{e}_{3}.
    \label{eq:reference_of_coordinate_system}
\end{equation}
Note also that if $\mb{X} = \mb{I}$, then $\bs{\xi} = \mring{\bs{\xi}}$. In other words, the identity element of the group relates to the origin of the coordinate system.

\subsection{Equivariance Condition and Right Input Action}
It is important to ensure that the propagation of the dynamics on the new manifold with group structure is diffeomorphic to the dynamics propagated on the original manifold. This requires defining an input transformation which encodes the equivariance of the system. To achieve this, the equivariance condition\;\eqref{eq:equivariant_system_condition} must be satisfied.

For this particular system, the left-side of \eqref{eq:equivariant_system_condition} is given by
\begin{equation}
\begin{split}
    D \phi_{\mb{X}}(\bs{\xi})[f(\bs{\xi}, \mb{u})] &= \lim_{t\to0} \frac{\phi_{\mb{X}}(\bs{\xi} + t f(\bs{\xi}, \mb{u})) - \phi_{\mb{X}}(\bs{\xi})}{t}\\
    &= -\mb{X}^{\top}\bs{S}(\bs{\omega})\bs{\xi} + \mb{X}^{\top}\mb{\bar{v}}.
\end{split}
\end{equation}
The right-side of the equality is given by
\begin{equation}
\begin{split}
    f(\phi_{\mb{X}}(\bs{\xi}), \psi_{\mb{X}}(\mb{u})) &= -\bs{S}(\psi^{\bs{\omega}}_{\mb{X}}(\mb{u}))\phi(\mb{X}, \bs{\xi}) + \psi_{\mb{X}}^{\mb{\bar{v}}}(\mb{u})\\    
    &= -\bs{S}(\psi^{\bs{\omega}}_{\mb{X}}(\mb{u}))\mb{X}^{\top}\bs{\xi}   +\psi_{\mb{X}}^{\mb{\bar{v}}}(\mb{u}).
    \label{eq:right_side_manifold_dynamics}
\end{split} 
\end{equation}
From here, we can conclude that a valid input mapping $\psi(\mb{X}, \mb{u})$ is given by
\begin{equation}
     \psi(\mb{X}, \mb{u}) \coloneqq (\psi^{\bs{\omega}}_{\mb{X}}(\mb{u}), \psi^{\mb{\bar{v}}}_{\mb{X}}(\mb{u})) = (\mb{X}^{\top}\bs{\omega},\,\mb{X}^{\top}\mb{\bar{v}}).
\label{eq:input_action}
\end{equation}
Replacing \eqref{eq:input_action} in \eqref{eq:right_side_manifold_dynamics} and using the property\;\eqref{eq:skew_rotation}, then
\begin{equation}
\begin{split}
    \hspace{-2mm}f(\phi_{\mb{X}}(\bs{\xi}), \psi_{\mb{X}}(\mb{u})) &= -\bs{S}(\mb{X}^{\top}\bs{\omega})\mb{X}^{\top}\bs{\xi} + \mb{X}^{\top}\mb{\bar{v}} \\
    &=-\mb{X}^{\top}\bs{S}(\bs{\omega})\bs{\xi} + \mb{X}^{\top} \mb{\bar{v}},
\end{split}
\end{equation}
which shows that, with this choice of right actions $\phi$ and $\psi$, the equivariance condition \eqref{eq:equivariant_system_condition} is satisfied.

\subsection{Equivariant System Lift onto the Group}
The equivariant lift of the system ${\Lambda(\bs{\xi},\mb{u}):\mc{M}\times \mc{U}\to\mf{g}}$ is a map that will be used to lift the dynamics of the system from the state manifold, with inputs on the input manifold, to the tangent space of the symmetry group\;\cite{mahony_observer_2022, MAHONY2021253}. The equivariant lift must satisfy the following two conditions\;\cite{mahony_equivariant_2020}:
\begin{align}
    D \phi_{\bs{\xi}}(\mb{X})_{\mid_{\mb{X}=\mb{I}}}[\Lambda(\bs{\xi},\mb{u})] &= f(\bs{\xi}, \mb{u}) ,
    \label{eq:lift_condition_1}\\
    \text{Ad}_{\mb{X}^{-1}}\Lambda(\bs{\xi},\mb{u}) &= \Lambda(\phi_{\mb{X}}(\bs{\xi}), \psi_{\mb{X}}(\mb{u})),
    \label{eq:lift_condition_2}    
\end{align}
where ${D \phi_{\bs{\xi}}(\mb{X}) : \mf{g} \to \mc{T}_{\bs{\xi}}\mc{M}}$. 
Condition \eqref{eq:lift_condition_1} is a necessary and sufficient condition to guarantee that the solutions of the lifted system project to solutions of the original system and condition \eqref{eq:lift_condition_2} ensures that the lift respects the symmetry\;\cite{mahony_observer_2022}.

To compute the equivariant lift of system \eqref{eq:example_system}, ${\Lambda(\bs{\xi}, \mb{u}):\mc{S}^2\times(\mbb{R}^3 \times \mbb{R}^3) \to\mf{so}(3)}$, we start by developing the expression of the first condition. The left side of \eqref{eq:lift_condition_1} is
\begin{equation}
\begin{split}
    D \phi_{\bs{\xi}}(\mb{X})[\Lambda(\bs{\xi},\mb{u})] &=\lim_{t\to0} \frac{\phi_{\bs{\xi}}(\mb{X}+t\Lambda(\bs{\xi},\mb{u})) - \phi_{\bs{\xi}}(\mb{X})}{t}\\
    &=\lim_{t\to0} \frac{(\mb{X}+t\Lambda(\bs{\xi},\mb{u}))^{\top}\bs{\xi} - \mb{X}^{\top}\bs{\xi}}{t}\\
    &=\Lambda(\bs{\xi},\mb{u})^{\top} \bs{\xi}=-\Lambda(\bs{\xi},\mb{u})\bs{\xi},
\end{split}
\end{equation}
where the property $\Lambda(\bs{\xi},\mb{u})^{\top} = -\Lambda(\bs{\xi},\mb{u})$ was applied, since this lift produces an output on the Lie-algebra, $\mf{so}(3)$. Making use of the right-side of \eqref{eq:lift_condition_1}, it can be concluded that
\begin{equation}
    -\Lambda(\bs{\xi},\mb{u})\bs{\xi} = -\bs{S}(\bs{\omega})\bs{\xi} + \mb{\bar{v}}.
    \label{eq:lift_pre_calculation_1}
\end{equation}
Taking the expression \eqref{eq:lift_pre_calculation_1} and using the fact that ${-\bs{S}(\mb{\bar{v}}\times\bs{\xi})\bs{\xi} = \bs{\Pi}_{\bs{\xi}}\mb{\bar{v}} = \mb{\bar{v}}}$, an equivariant lift of the system is given by
\begin{equation}
    \Lambda(\bs{\xi},\mb{u}) = \bs{S}(\bs{\omega} + \mb{\bar{v}} \times \bs{\xi}),
    \label{eq:lift_expression}
\end{equation}

Given the expression for the equivariant lift, we must now verify that the two conditions given by \eqref{eq:lift_condition_1} and \eqref{eq:lift_condition_2} are satisfied. Starting with the first condition, we substitute the proposed lift in \eqref{eq:lift_pre_calculation_1}, which gives
\begin{equation}
\begin{split}
    -\Lambda(\bs{\xi},\mb{u}) \bs{\xi} &=-\bs{S}(\bs{\omega} + \mb{\bar{v}} \times \bs{\xi}) \bs{\xi} \\
    &=-\bs{S}(\bs{\omega}) \bs{\xi} - \bs{S}(\mb{\bar{v}} \times \bs{\xi}) \bs{\xi}.
\end{split}
\end{equation}
Using properties of the skew-map and the cross-product, then
\begin{equation}
    -\Lambda(\bs{\xi},\mb{u}) \bs{\xi}=-\bs{S}(\bs{\omega})\bs{\xi} + \mb{\bar{v}}.
\end{equation}
Thus, we conclude that the first condition of the equivariant lift is satisfied. Let us now analyze the second condition. Making use of the property \eqref{eq:skew_rotation}, the left-side of \eqref{eq:lift_condition_2} becomes
\begin{equation}
\begin{split}
    \text{Ad}_{\mb{X}^{-1}}\Lambda(\bs{\xi}, \mb{u}) &=\mb{X}^{\top} \bs{S}(\bs{\omega} + \mb{\bar{v}} \times \bs{\xi}) \mb{X} \\
    &=\bs{S}\left(\mb{X}^{\top} (\bs{\omega} + \mb{\bar{v}} \times \bs{\xi})\right).
\end{split}
\end{equation}
Expanding on the right-side of \eqref{eq:lift_condition_2} yields
\begin{equation}
\begin{split}
    \Lambda(\phi_{\mb{X}}(\bs{\xi}), \psi_{\mb{X}}(\mb{u}))&=\bs{S}(\mb{X}^{\top}\bs{\omega} + \mb{X}^{\top}\mb{\bar{v}} \times \mb{X}^{\top}\bs{\xi})\\
    &=\bs{S}\left(\mb{X}^{\top} (\bs{\omega} + \mb{\bar{v}} \times \bs{\xi})\right).
\end{split}
\end{equation}
Thus, both conditions are satisfied and we conclude that the proposed lift is in fact an equivariant lift of the system. An intuition for the lift expression is presented in Appendix\,\ref{app:lift_analysis}.

\subsection{Dynamics of the Lifted System}
With the lift operation defined, the dynamics of the system can be propagated on the matrix Lie group, according to
\begin{equation}
    \dot{\mb{X}} = \mb{X} \Lambda(\phi(\mb{X}, \mathring{\bs{\xi}}), \mb{u}).
    \label{eq:lifted_group_dynamics}
\end{equation}
For this particular system, the dynamics on the Lie group are given by
\begin{equation}
    \dot{\mb{X}} = \mb{X} \bs{S}(\bs{\omega} + \mb{\bar{v}} \times \bs{\xi}).
    \label{eq:lifted_group_dynamics2}
\end{equation}

\section{Equivariant observer design} 
\label{sec:equivariant_observer_design}

In this section the observer structure is introduced. We start by defining the estimation error on the Lie group and make use of the group actions $\phi$ and $\psi$, and the equivariant lift $\Lambda$ derived in the previous section to propose a correction term that stabilizes the system. The derived observer is also expressed back in the original manifold.

\subsection{Observer on the Group}
\begin{figure}
	\centering
    \includegraphics[width=\columnwidth]{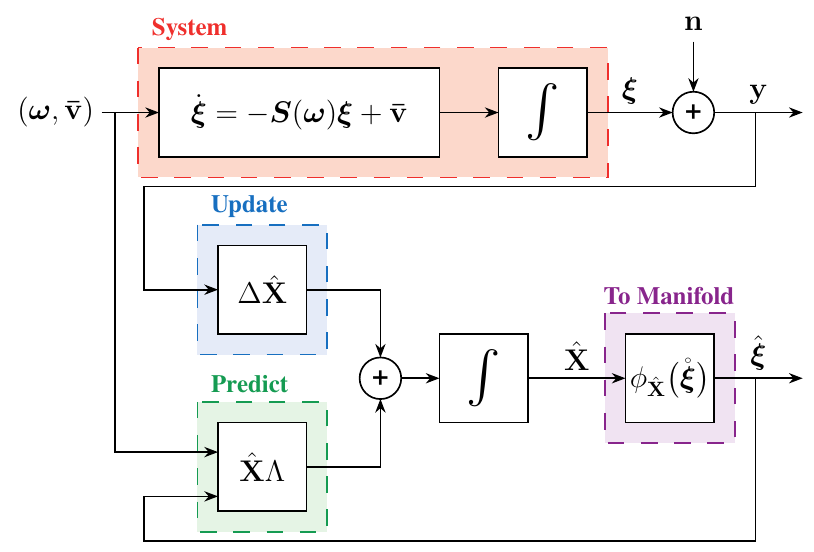}
	\caption{Observer design on the Lie group.}
	\label{fig:complementary_filter_diagram}
\end{figure}

The observer state on the Lie group is denoted by $\hat{\mb{X}}$. The equivariant observer is given by 
\begin{equation}
    \dot{\hat{\mb{X}}} = \hat{\mb{X}}\Lambda(\phi_{\mathring{\bs{\xi}}}(\hat{\mb{X}}), \mb{u}) + \Delta\hat{\mb{X}},
    \label{eq:observer_definition}
\end{equation}
where ${\hat{\mb{X}}\Lambda(\phi_{\mathring{\bs{\xi}}}(\hat{\mb{X}}), \mb{u})}$ is a replica of the lifted system's dynamics and ${\Delta:G\times\mc{Y}\times\mc{U}\to\mf{g}}$ is a correction term, following to the structure proposed in\;\cite{mahony_observer_2022}. In Fig.\;\ref{fig:complementary_filter_diagram}, we present the diagram of the system and observer, separated in a prediction part, which uses the replica of the lifted system, and an update part, which uses the correction term.

The state estimate of the observer can be expressed in the state manifold $\mc{M}$, by direct application of \eqref{eq:reference_of_coordinate_system}, according to
\begin{equation}
    \hat{\bs{\xi}} := \phi(\hat{\mb{X}}, \mring{\bs{\xi}}) = \hat{\mb{X}}^{\top}\mb{e}_3.
    \label{eq:estimate_in_original_manifold}
\end{equation}
 
\subsection{Error System}

The global state error, defined on the manifold, is given by ${\mb{e} = \phi(\hat{\mb{X}}^{-1}, \bs{\xi}) = \phi(\hat{\mb{X}}^{\top}, \bs{\xi}) \in \mc{M}}$. The error can be further expressed as
\begin{equation}
    \mb{e} = \phi(\hat{\mb{X}}^{\top}, \bs{\xi}) = \phi(\hat{\mb{X}}^{\top}, \phi(\mb{X}, \mring{\bs{\xi}})) = \phi(\mb{X}\hat{\mb{X}}^{\top}, \mring{\bs{\xi}}).
    \label{eq:global_state_error_definition}
\end{equation}
We define the error on the Lie group as ${\mb{E} = \mb{X}\hat{\mb{X}}^{\top} \in SO(3)}$, such that ${\mb{e} = \phi(\mb{E}, \mring{\bs{\xi}})}$. The goal of the observer is to drive the error $\mb{e}$ to the origin $\mring{\bs{\xi}}$ or, equivalently, to drive the error $\mb{E}$ to the stabilizer subgroup ${\tr{stab}_{\phi}(\mring{\bs{\xi}})}$.
The time-derivative of the error in the group is given by
\begin{equation}
\begin{split}
    \dot{\mb{E}}&= \dot{\mb{X}} \hat{\mb{X}}^{\top} + \mb{X}\dot{\hat{\mb{X}}}^{\top}\\ 
                &= \mb{X} \big(\Lambda(\phi_{\mring{\xi}}(\mb{X}),\mb{u}) - \Lambda(\phi_{\mring{\xi}}(\hat{\mb{X}}),\mb{u})\big) \hat{\mb{X}}^{\top} - \mb{X}\hat{\mb{X}}^{\top}\Delta,
\end{split}
\label{eq:group_error_derivative_pre}
\end{equation}
where the fact that ${\Delta \in \mf{so}(3)}$ and ${\Delta^{\top} = -\Delta}$ was used. Taking into account that ${\mb{E}=\mb{X}\hat{\mb{X}}^{\T}}$ and ${\hat{\mb{X}}^{\T}\hat{\mb{X}}=\mb{I}}$, and using the equivariant condition \eqref{eq:lift_condition_2}, the above expression can be written as
\begin{equation}
\begin{split}
    \dot{\mb{E}} &= \mb{E}\big(\tr{Ad}_{\hat{\mb{X}}}\Lambda(\phi_{\mring{\xi}}(\mb{X}),\mb{u}) - \tr{Ad}_{\hat{\mb{X}}}\Lambda(\phi_{\mring{\xi}}(\hat{\mb{X}}),\mb{u}) - \Delta\big)\\
                &= \mb{E}\big(\Lambda(\phi_{\mring{\xi}}(\mb{E}),\mring{\mb{u}}) - \Lambda(\mring{\xi},\mring{\mb{u}}) - \Delta\big),
\end{split}
\label{eq:group_error_derivative}
\end{equation}
where ${\mring{\mb{u}} = (\mring{\bs{\omega}},\mring{\bar{\mb{v}}}) = \psi_{\hat{\mb{X}}^{\top}}(\mb{u})}$. The time-derivative of the error can still be further developed, yielding
\begin{equation}
    \dot{\mb{E}}=  \mb{E}\big(\bs{S}(\mring{\bar{\mb{v}}}\times\mb{E}^{\top}\mring{\bs{\xi}} - \mring{\bar{\mb{v}}}\times\mring{\bs{\xi}}) -\Delta\big).
    \label{eq:group_error_derivative_final}
\end{equation}

\subsection{Correction Term Design}
The next step is to design a correction term $\Delta$ such that the estimate $\hat{\bs{\xi}}$ converges to the true bearing $\bs{\xi}$. 

\begin{theorem*}
    Consider the lifted system described by \eqref{eq:lifted_group_dynamics2} and the error dynamics given by \eqref{eq:group_error_derivative_final}. Taking the correction term 
    \begin{equation}
        \Delta = \bs{S}(\hat{\mb{X}}\mb{\bar{v}}\times\hat{\mb{X}}\mb{y} - \hat{\mb{X}}\mb{\bar{v}}\times\mring{\bs{\xi}} + k\hat{\mb{X}}\mb{y} \times \mring{\bs{\xi}}),
        \label{eq:correction_term}
    \end{equation}
    with gain ${k>0}$, the observer given by \eqref{eq:observer_definition} is almost globally asymptotically stable.
\end{theorem*}

\begin{proof}
    First, consider the Lyapunov function $V(\mb{E})$, given by
    \begin{equation}
        V(\mb{E}) := \frac{1}{2} \norm{\mb{E}^{\top}\mring{\bs{\xi}}-\mring{\bs{\xi}}}^2 = 1 - \mring{\bs{\xi}}^{\top}\mb{E}\,\mring{\bs{\xi}}.
        \label{eq:lyapunov_function}
    \end{equation}
    Taking its time-derivative and using \eqref{eq:group_error_derivative} gives 
    \begin{equation}
        \dot{V}(\mb{E}) = -\mring{\bs{\xi}}^{\top}\mb{E}\big(\Lambda(\phi_{\mring{\xi}}(\mb{E}),\mring{\mb{u}}) - \Lambda(\mring{\xi},\mring{\mb{u}}) - \Delta\big)\mring{\bs{\xi}}.
        \label{eq:time_derivative_lyapunov_function}
    \end{equation}
    Then, note that ${\hat{\mb{X}}\mb{y}=\mb{E}^{\top}\mring{\bs{\xi}}}$ and ${\hat{\mb{X}}\mb{\bar{v}}=\mring{\bar{\mb{v}}}}$. As such, the correction term \eqref{eq:correction_term} can be expressed as
    \begin{equation}
    \begin{split}
        \Delta  &= \bs{S}(\mring{\bar{\mb{v}}}\times\mb{E}^{\top}\mring{\bs{\xi}} - \mring{\bar{\mb{v}}}\times\mring{\bs{\xi}} + k\mb{E}^{\top}\mring{\bs{\xi}} \times \mring{\bs{\xi}})\\
                &= \Lambda(\phi_{\mring{\bs{\xi}}}(\mb{E}), \mring{\mb{u}}) - \Lambda(\mring{\bs{\xi}}, \mring{\mb{u}}) + k\bs{S}(\mb{E}^{\top}\mring{\bs{\xi}} \times \mring{\bs{\xi}}).
    \end{split}
    \label{eq:correction_term_2}
    \end{equation}
    Taking the Lyapunov function \eqref{eq:lyapunov_function} and replacing the correction term \eqref{eq:correction_term_2} in its time-derivative \eqref{eq:time_derivative_lyapunov_function} yields
    \begin{equation}
        \begin{split}
            \dot{V}(\mb{E}) &= -\mring{\bs{\xi}}^{\top}\mb{E}\,\big(-k\bs{S}(\mb{E}^{\top}\mring{\bs{\xi}} \times \mring{\bs{\xi}}) \big)\mring{\bs{\xi}}\\
            &= -k\norm{\mb{E}^{\top}\mring{\bs{\xi}} \times \mring{\bs{\xi}} }^2 \leq 0.
        \end{split}
        \label{eq:lyap_function_derivative_2}
    \end{equation}
    The set of points where ${V(\mb{E})=0}$ is the stabilizer ${\tr{stab}_{\phi}(\mring{\bs{\xi}})}$ and the set of points where ${\dot{V}(\mb{E})=0}$ is ${\{\mb{E} \in SO(3) \mid \mb{E}^{\top}\mring{\bs{\xi}} = \pm \mring{\bs{\xi}}\}}$, which corresponds either to matrices on the stabilizer subgroup or to rotations of $180^{\circ}$ about any axis orthogonal to $\mring{\bs{\xi}}$. The latter are undesired equilibria that are unstable, as shown in\;\cite{CUNHA20081013,BHAT200063}. Therefore, the largest invariant set where ${\dot{V}(\mb{E})=0}$ is the ${\tr{stab}_{\phi}(\mring{\bs{\xi}})}$. 

    Hence, since the group error $\mb{E}$ converges to the ${\tr{stab}_{\phi}(\mring{\bs{\xi}})}$ and the global state error $\mb{e}$ converges to the origin $\mring{\bs{\xi}}$, the observer is almost globally asymptotically stable.
\end{proof}

\subsection{Observer on the Manifold}
\label{subsec:observer_on_manifold}

The derived equivariant observer can also be expressed in the state manifold, i.e. the unit-sphere, using the group action $\phi$\;\cite{MAHONY2021253}. Using the fact that the time-derivative of \eqref{eq:estimate_in_original_manifold} is given by $\dot{\hat{\bs{\xi}}} = \dot{\hat{\mb{X}}}^{\top} \mring{\bs{\xi}}$, and the observer definition \eqref{eq:observer_definition}, the observer dynamics on $\mc{M}$ can be given by
\begin{equation}
    \dot{\hat{\bs{\xi}}} = (\hat{\mb{X}} \Lambda(\phi_{\mring{\bs{\xi}}}(\hat{\mb{X}}), \mb{u}) + \Delta \hat{\mb{X}})^{\top}\mring{\bs{\xi}}.
\label{eq:observer_manifold_intermediate}
\end{equation}
Replacing \eqref{eq:lift_expression}, \eqref{eq:estimate_in_original_manifold} and \eqref{eq:correction_term}, in \eqref{eq:observer_manifold_intermediate} yields
\begin{equation}
\dot{\hat{\bs{\xi}}} = -\mb{S}(\bs{\omega} + \mb{\bar{v}}\times\mb{y}) \hat{\bs{\xi}} + k \bs{\Pi}_{\hat{\bs{\xi}}}\mb{y}.
\label{eq:observer_manifold}
\end{equation}
For intuition, note that the correction ${k \bs{\Pi}_{\hat{\bs{\xi}}}\mb{y}=k \bs{\Pi}_{\hat{\bs{\xi}}}(\mb{y}-\hat{\bs{\xi}})}$, due to the properties of the projection operator.
Observe also that the bearing system \eqref{eq:example_system} can be expressed as
\begin{equation}
    \dot{\bs{\xi}} = -\mb{S}(\bs{\omega} + \mb{\bar{v}}\times\bs{\xi}) \bs{\xi}.
    \label{eq:example_system_alternative}
\end{equation}
By comparing the observer on the manifold \eqref{eq:observer_manifold} and the alternative expression for the system \eqref{eq:example_system_alternative}, we see that the equivariant observer is subtly different from a typical observer, which usually consists of a replica of the system and a correction term. While the observer on the group had such structure, with a replica of the lifted system, notice that the observer on the manifold incorporates the measurement $\mb{y}$ directly on what would be the replica of the system on the manifold, i.e., ${-\mb{S}(\bs{\omega} + \mb{\bar{v}}\times\mb{y}) \hat{\bs{\xi}}}$ instead of ${-\mb{S}(\bs{\omega} + \mb{\bar{v}}\times\hat{\bs{\xi}}) \hat{\bs{\xi}}}$. This difference is crucial to obtaining the almost global asymptotical stability property. 



\section{Numerical Results}
\label{sec:numerical_results}

To evaluate the performance of the observer designed in Section\;\ref{sec:equivariant_observer_design}, we conduct numerical simulations in MATLAB\textsuperscript{\tiny\textregistered}. 
Let us define an auxiliary linear velocity term, ${\mb{\bar{v}^{\prime}}(t)}$, such that ${\mb{\bar{v}}(t) = \bs{\Pi}_{\mb{b}(t)}\mb{\bar{v}^{\prime}(t)}}$.
We assume that the linear and angular velocities, ${\mb{\bar{v}^{\prime}}(t)\in\mbb{R}^{3}}$ and ${\bs{\omega}(t)\in\mbb{R}^{3}}$, have sinusoidal components given by ${\mr{\bar{v}^{\prime}}_i(t) = A_{i}^{\mb{v}}\sin(2\pi\nu_{i}^{\mb{v}} t+\varphi_{i}^{\mb{v}})}$ and ${\omega_i(t)=A_{i}^{\bs{\omega}}\sin(2\pi\nu_{i}^{\bs{\omega}} t+\varphi_{i}^{\bs{\omega}})}$, respectively, for ${i\in\{1,2,3\}}$. 
The amplitudes, frequencies, and phases are set randomly, with 
${A_{i}^{\mb{v}},A_{i}^{\bs{\omega}},\nu_{i}^{\mb{v}},\nu_{i}^{\bs{\omega}}\sim\mathpzc{U}([0,10])}$, and ${\varphi_{i}^{\mb{v}},\varphi_{i}^{\bs{\omega}}\sim\mathpzc{U}(-\pi, \pi)}$, where ${\mathpzc{U}([a,b])}$ denotes a uniform distribution in the interval ${[a,b]}$.
We further assume that both inputs are corrupted by additive white Gaussian noise following the distribution ${\mathpzc{N}(\mb{0},0.1^{2}\mb{I})}$.
The bearing measurements are affected by rotation noise with an angle standard deviation of $5^{\circ}$ and include outliers occurring with a probability of \SI{1}{\percent}.
The initial value of the system, $\bs{\xi}(0)$, is generated randomly, uniformly on $\mc{S}^{2}$, the initial filter group state is ${\hat{\mb{X}}(0) = \mb{I}}$, and the gain is set to $k=1$. 

\begin{figure}
    \centering
    \includegraphics[width=\columnwidth]{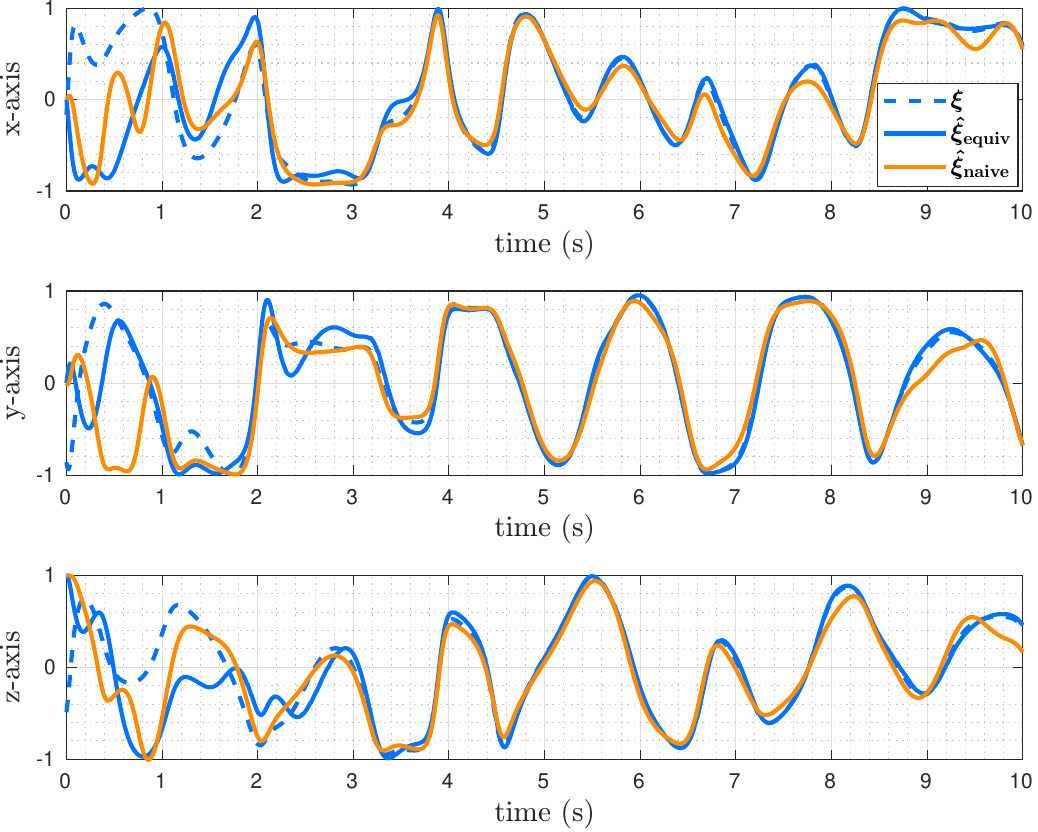}
    \caption{Representative simulation with inputs and measurements corrupted by noise. The true bearing is depicted in dashed blue lines, while the estimates provided by the equivariant and naive observers are provided in solid blue and orange lines, respectively.}
    \label{fig:simulation_observer_estimate}
\end{figure}

\begin{figure}
    \centering
    \begin{subfigure}{0.49\textwidth}
        \centering
        \includegraphics[width=\linewidth]{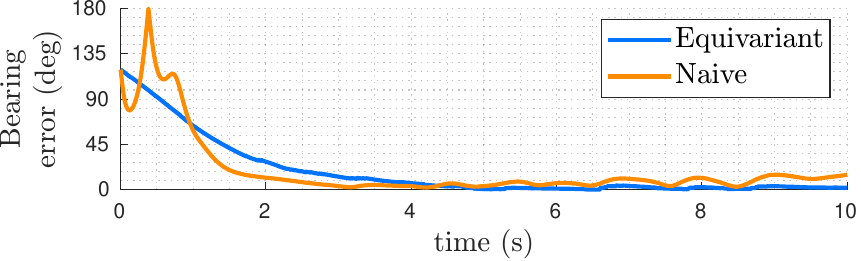}
        \caption{Error between the true and estimated bearing. Equivariant observer error depicted in blue and the naive observer depicted in orange.}
        \label{fig:simulation_bearing_error}
    \end{subfigure}
    \hfill
    \begin{subfigure}{0.49\textwidth}
        \centering
        \includegraphics[width=\linewidth]{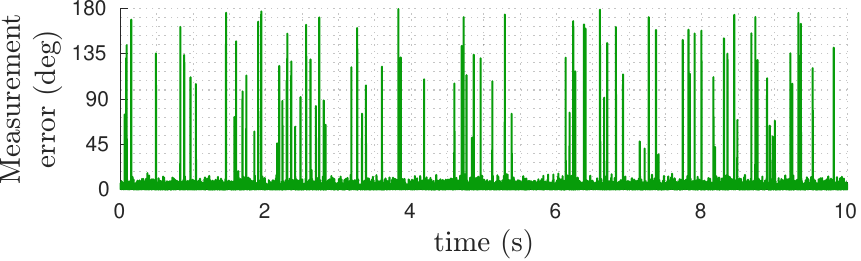}
        \caption{Error between the true and measured bearing.}
        \label{fig:simulation_measurement_error}
    \end{subfigure}
    \caption{Evolution of the estimation error and measurement outliers.}
    \label{fig:simulation_comparison}
\end{figure}

We also compare the equivariant observer to a more naive approach alluded to in \ref{subsec:observer_on_manifold}, given by a replica of the system with a correction term, according to
\begin{equation}
    \dot{\hat{\bs{\xi}}} =  -\mb{S}(\bs{\omega} + \mb{\bar{v}}\times\hat{\bs{\xi}}) \hat{\bs{\xi}} + k\bs{\Pi}_{\hat{\bs{\xi}}} \mb{y}.
    \label{eq:naive_observer_manifold}
\end{equation}
In Fig.\;\ref{fig:simulation_observer_estimate}, we present the evolution of the bearing estimation in a representative simulation. In Fig.\;\ref{fig:simulation_bearing_error}, we display the angle error between the true bearing, $\bs{\xi}$, and the estimated bearing, $\hat{\bs{\xi}}$, given by ${\arccos{\big(\hat{\bs{\xi}}^{\top}\bs{\xi}\big)}}$. 
The plot shows the angle error for both the equivariant observer (in blue) and the naive observer (in orange). 
In Fig.\;\ref{fig:simulation_measurement_error}, we present the angle error between the true and measured bearings, depicting the error of outlier measurements.

As can be seen from the plots, even though the inputs and the measurements were corrupted by noise, the equivariant observer accurately estimates the bearing, converging smoothly to the true value. The naive version of the observer on the manifold shows an irregular behavior and, though it seems to converge at approximately the same time as the equivariant observer, the estimates then diverge due to the noise.

\section{Conclusion}\label{sec:conclusion}

In this paper, an equivariant observer for a bearing system based on the dynamics of features in an image plane was derived. This work expands on the well-established problem of devising an equivariant observer for bearing estimation to a system that includes both linear and angular velocity inputs. The necessary maps were defined according to the underlying group symmetry, which allowed us to derive an equivariant lift and design a correction term that made the observer almost globally asymptotically stable.
The observer was expressed back on the manifold and analyzed. The method was tested in simulation and the results were presented.

\appendices


\section{Proof that $\phi$ is a transitive right group action}
\label{app:right_action_phi_proof}
\begin{proof}
To be a right group action, $\phi: G \times \mc{M} \to \mc{M}$ must verify $\phi(\mb{X}\mb{Y}, \bs{\xi}) = \phi(\mb{Y}, \phi(\mb{X}, \bs{\xi}))$. The left side of this expression is given by $\phi(\mb{X}\mb{Y}, \bs{\xi}) = (\mb{X}\mb{Y})^\T \,\bs{\xi}$. Developing the right side of the equality yields $\phi(\mb{Y}, \phi(\mb{X}, \bs{\xi})) = \mb{Y}^\T(\mb{X}^\T\,\bs{\xi}) = (\mb{X}\mb{Y})^\T\bs{\xi}$. 
To be a transitive group action, then for all $\bs{\xi}, \bs{\xi}' \in \mc{S}^2$, there is an $\mb{X}\in SO(3)$, such that $\phi(\mb{X}, \bs{\xi}) = \mb{X}^{\top} \bs{\xi} = \bs{\xi}'$, i.e., $\bs{\xi}'$ is the result of rotating $\bs{\xi}$ by some rotation matrix $\mb{X}^{\top}$. This rotation can be described by an axis of rotation given by the cross product $\bs{\xi}\times \bs{\xi}'$ and an angle equal to the angle between $\bs{\xi}$ and $\bs{\xi}'$. The matrix $\mb{X}^{\top}$ encodes this rotation.
\end{proof}
\section{Analysis of the proposed equivariant lift}
\label{app:lift_analysis}

To gain intuition for the lift of system \eqref{eq:example_system}, we start by analyzing the well studied simplified system ${\dot{\bs{\xi}} = -\bs{S}(\bs{\omega}) \bs{\xi}}$. For this system, it is known that the lifted dynamical system is ${\dot{\mb{X}} = \mb{X} \bs{S}(\bs{\omega})}$, i.e. the system lift is ${\Lambda(\bs{\xi}, \bs{\omega}) = \bs{S}(\bs{\omega})}$\;\cite{mahony_observer_2022}. Note that, in system \eqref{eq:example_system}, $\mb{\bar{v}}$ lives in the tangent plane of the bearing and can be thought of as inducing an angular velocity ${\bs{\Omega} \in \mbb{R}^{3}}$ about the bearing. Hence, let us equate both dynamics
\begin{equation}
     -\bs{\Omega} \times \bs{\xi} = \mb{\bar{v}}.
\end{equation}
By applying the cross product of $\bs{\xi}$ to both sides of the equation, we get
\begin{equation}
    -(\bs{\Omega} \times \bs{\xi})\times \bs{\xi} = \mb{\bar{v}}\times \bs{\xi}.
    \label{eq:appendix_cross_product_of_induced_velocity}
\end{equation}
Further developing, the left-side of the equality becomes
\begin{equation}
    -(\bs{\Omega} \times \bs{\xi})\times \bs{\xi} = \bs{\Omega} - (\bs{\xi}\bs{\xi}^{\top})\bs{\Omega} = \bs{\Pi}_{\bs{\xi}}\bs{\Omega}.
    \label{eq:appendix_skew_to_proj}
\end{equation}
Decomposing the angular velocity ${\bs{\Omega} = \bs{\Omega}_{\perp} + \bs{\Omega}_{\|}}$ in components orthogonal and parallel to the bearing vector, respectively, and replacing \eqref{eq:appendix_skew_to_proj} in \eqref{eq:appendix_cross_product_of_induced_velocity} yields
\begin{equation}
    \bs{\Pi}_{\bs{\xi}}(\bs{\Omega}_{\perp} + \bs{\Omega}_{\|}) = \bs{\Omega}_{\perp}  = \mb{\bar{v}}\times \bs{\xi}.
\end{equation}
We conclude that $\bs{\Omega}_{\perp}$ is the component generated by $\mb{\bar{v}}$. As such, the lift is ${\Lambda(\bs{\xi}, \mb{u}) = \bs{S}(\bs{\omega} + \bs{\Omega}_{\perp}) = \bs{S}(\bs{\omega} + \mb{\bar{v}}\times \bs{\xi})}$. 

\bibliographystyle{IEEEtran}
\bibliography{bibliography}

\end{document}